\documentclass{article}
\usepackage{spconf,amsmath,graphicx,hyperref}
\usepackage[]{adjustbox}
\usepackage{bm} 
\usepackage{tabularx}   
\usepackage{balance}

\title{Audiocards: Structured metadata improves audio language models for sound design}

\name{
Sripathi Sridhar $^{1,2}$,
      Prem Seetharaman$^{2}$,
      Oriol Nieto$^{2}$,
      Mark Cartwright$^{1}$,
      Justin Salamon$^{2}$
\address{$^{1}$New Jersey Institute of Technology \;
$^{2}$Adobe Research \\
}
}

\begin{document}
\ninept
\maketitle

\begin{figure*}[t]
  \centering
  {\includegraphics[width=\textwidth]{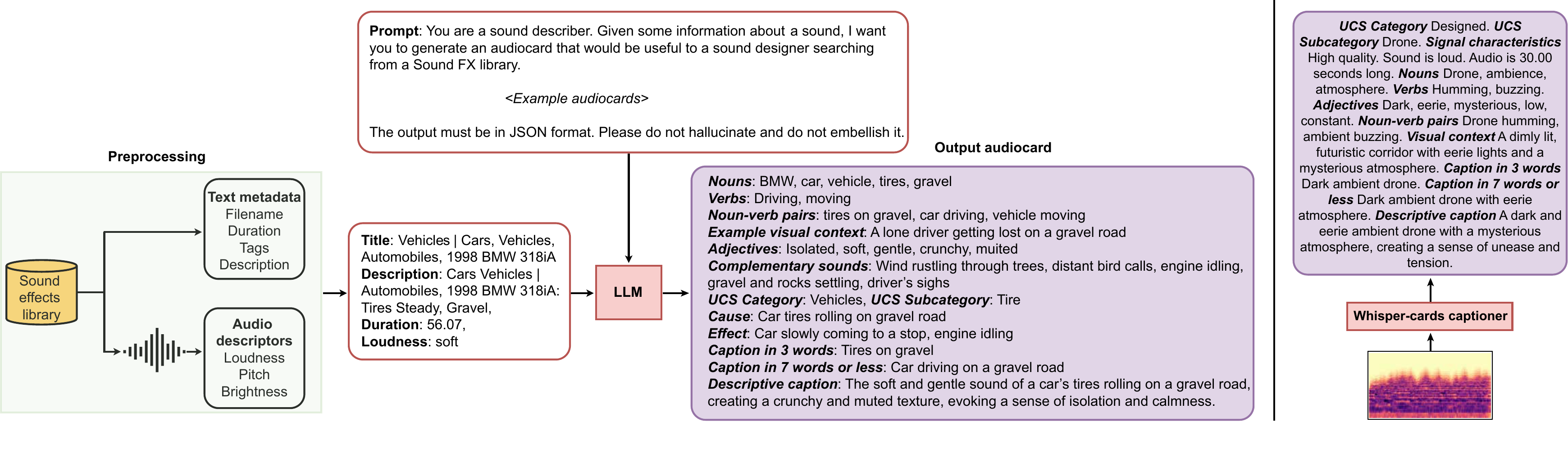}}
  \caption{Left: We propose \textit{Audiocards}, structured metadata which describes an audio file with attributes relevant to sound designers. We prompt an LLM with the available text metadata and audio descriptors, and generate an audiocard, which can be used for text-based search and to train audio-language models. Right: Audiocard generated by our Whisper-cards audio captioner from input audio without text metadata.}
  \label{fig:audiocard_generation}
\end{figure*}

\begin{abstract}
Sound designers search for sounds in large sound effects libaries using aspects such as sound class or visual context. However, the metadata needed for such search is often missing or incomplete, and requires significant manual effort to add. Existing solutions to automate this task by generating metadata, i.e. captioning, and search using learned embeddings, i.e. text-audio retrieval, are not trained on metadata with the structure and information pertinent to sound design. To this end we propose audiocards, structured metadata grounded in acoustic attributes and sonic descriptors, by exploiting the world knowledge of LLMs. We show that training on audiocards improves downstream text-audio retrieval, descriptive captioning, and metadata generation on professional sound effects libraries. Moreover, audiocards also improve performance on general audio captioning and retrieval over the baseline single-sentence captioning approach. We release a curated dataset of sound effects audiocards to invite further research in audio language modeling for sound design.

\end{abstract}


\section{Introduction}
Sound designers need to search in large sound libraries to find relevant sounds for their tasks. To achieve this, they typically search using aspects such as sound effect class, visual context (e.g. ``car chase scene"), acoustic attributes, and causal relationships \cite{weck2024language, sonnenschein2001sound}. However, existing professional sound libraries do not always have this information in their metadata, and significant human effort is required to provide this information, which is especially challenging for personal sound effects libraries recorded and/or collected by sound designers \cite{noauthor_frank_nodate, forshee_top_nodate}.

Audio captioning models can predict descriptive information about audio \cite{wu_improving_2024, mei_wavcaps_2024, zhu_cacophony_2024}, but they are generally trained on sentence-long captions from crowd-sourced workers \cite{drossos_clotho_2020}, which do not necessarily contain the attributes or structure relevant to sound designers. While text-audio retrieval models can search for sounds without metadata using just a text query \cite{wu2023large, elizalde2023clap}, these embeddings are also trained on similar sentence captions which are not aligned with the shorter queries favored by users \cite{weck2024language}. Furthermore, sound designers prefer to assess search results without having to listen to each file, which motivates the need for descriptive metadata \cite{liu2025exploring}.

To address these gaps, we propose structured metadata called \textit{audiocards}. Audiocards are multi-field descriptions of audio files grounded in their acoustic attributes, sound classes, and usage context. Based on conversations with a professional sound designer, we include fields such as nouns, noun-verb pairs, and Universal Category System (UCS) \cite{noauthor_universal_nodate} tags, as well as visual context and adjectives, which are aligned with sound designer needs \cite{weck2024language, sonnenschein2001sound, riege_tta_2022}. To construct an audiocard, we few-shot prompt an off-the-shelf LLM with metadata about an audio file, and a set of example handwritten cards, as seen in Figure \ref{fig:audiocard_generation}. Generating captions after all other audiocard fields also leads to more descriptive captions by grounding in attributes relevant to sound design and inducing chain-of-thought like reasoning \cite{wei2022chain}.

With audiocards, we can support both metadata generation and text-audio retrieval for sound design in multiple use cases. When text metadata is available, audiocards generated from LLMs can be used for keyword search on sound-design relevant fields. However, we can also train downstream captioning models on audiocards, enabling structured metadata and descriptive caption generation for sound libraries with incomplete or missing metadata. Finally, we can train text-audio embedding models on audio and audiocard field pairs to enable embedding-based audio retrieval aligned to the needs of sound designers. Overall, these approaches support sound design search scenarios such as using noun-verb pairs, e.g. ``dog barking'' \cite{weck2024language} or an example visual scene \cite{sonnenschein2001sound} e.g. ``haunted house", and they can generate sound-design-relevant fields and descriptions that can be used to organize sound effects libraries and provide additional context during retrieval. 

To validate the usefulness of audiocards we curate a human-verified subset of a professional sound effects library Audition SFx, called ASFx eval. We evaluate our captioning model on metadata generation and a descriptive captioning task on ASFx eval, as well as general audio captioning \cite{drossos_clotho_2020}. We evaluate our text-audio representation model on retrieval, both on a professional sound effects library and a general retrieval dataset \cite{drossos_clotho_2020}. We compare against state-of-the-art large audio language models (LALMs) and general contrastive text-audio embedding models on their respective tasks. 

Our experiments show that audiocards improve audio-language modeling for sound design, specifically, that (1) we can generate audiocards from audio without accompanying metadata better than state-of-the-art LALMs; (2) training on audiocards leads to significantly better captions for professional sound effects libraries compared to both a baseline approach and state-of-the-art LALMs; and (3) training on audiocards improves embedding-based text-audio retrieval when using human-annotated captions as queries on a professional sound effects library.
\vspace{-0.5em}

\section{Audiocards}
\vspace{-0.5em}
\label{sec:audiocards}
We propose \textit{audiocards}, structured descriptions of audio files with fields consisting of attributes contextually relevant to users' audio understanding activity and goals. In this work, we focus on audiocards to aid in audio-language modeling for sound design.


\subsection{Fields}
\vspace{-0.5em}
We propose audiocard fields for sound design based on conversations with a professional sound designer. We opt for a noun-verb structure to describe the key sound-producing actors (what) and actions (how) in the audio file \cite{wijngaard2023aces}, through the fields \textit{nouns, verbs} and \textit{noun-verb pairs}. We also include the \textit{Universal Category System (UCS) category} and \textit{UCS subcategory} \cite{noauthor_universal_nodate}, which is a hierarchical ontology consisting of 82 categories and 753 sub-categories, commonly used to tag sound effects in professional sound libraries. This helps ground the audiocard vocabulary in what professional sound designers may use.

To better exploit the world knowledge of the LLM, we include a field called \textit{example visual context} that acts as a bridge to potential visual scenes where this sound effect may be used. We provide additional auditory context using the fields of \textit{complementary sounds}, which denotes potential co-occurring sounds, and \textit{cause, effect} which aim to exploit a physical understanding of how the sound was created. We include an \textit{adjectives} field that allows scope for sonic and emotional descriptive keywords. Conditioned on these fields, we then include \textit{caption in 3 words}, \textit{caption in 7 words or less}, and \textit{descriptive caption}, to round out the card. A full example audiocard is depicted in Figure \ref{fig:audiocard_generation}.

To improve the auditory grounding of these cards and mitigate hallucinations, we compute several descriptors from the audio signal. Specifically, we include duration, loudness, brightness, and pitch at the top of the audiocards. We pick these audio descriptors based on prior work that appended audio descriptors in captions to improve text-audio generation \cite{kumar_sila_2024}. We include duration in seconds directly from the audio metadata. We compute loudness and bin them as \textit{very soft} (-70 LKFS to -55 LKFS), \textit{soft} (-55 LKFS to -40 LKFS), \textit{loud} (-30 LKFS to -15 LKFS), and \textit{very loud} (greater than -15 LKFS). We compute the spectral centroid as brightness, and classify values as \textit{dull} (spectral centroid $<$ 45) or \textit{bright} (spectral centroid $>$ 65) based on the distribution of values for our training data. We compute pitch using CREPE \cite{kim_crepe_2018} on normalized and segmented audio, and classify pitch as \textit{low} (pitch $<$ 1.5 octaves) and \textit{high} (pitch $>$ 3.5 octaves). Values that fall outside these ranges are mapped to an empty string, allowing the model to learn from natural language audio descriptions with low ambiguity.

\vspace{-0.75em}
\subsection{Training Data}
\vspace{-0.5em}
We gather a large mix of proprietary, licensed sound effect datasets and publicly available CC-licensed general audio datasets, consisting of approximately 2M audio samples with corresponding metadata. 

\vspace{-0.75em}
\subsection{Human verified sound effects evaluation dataset}
\vspace{-0.5em}
Audio language models are typically evaluated on audio captioning datasets like Clotho \cite{drossos_clotho_2020}. However, these datasets may not be the best indicators of model performance in a sound design setting, as they do not reflect the data distribution of a sound effects database. Hence, we include Adobe Audition SFx, a publicly available library of $~$10k professionally recorded sound effects in our model evaluations \footnote{https://www.adobe.com/products/audition/offers/adobeauditiondlcsfx.html}. 

Audition SFx primarily contains individually recorded sounds, and a small percentage of ambience recordings with overlapping sources. There are sound effects from 22 categories, such as Transportation and Production Elements, each with a descriptive title such as \texttt{Drones/Drone Production Element Imaging Deep Synth Rumble Atonal 01.wav}. We synthesized audiocards from these titles that we then used to evaluate our models. We extracted an evaluation subset for which we manually verified the generated audiocards accurately describe the audio file without incoherence or hallucination issues.

We built an annotation interface and annotated random examples on whether to keep them in the evaluation subset by listening to the audio file and examining the corresponding audiocard. We filtered out sound effects from the Noise, Tones, DTMF, DC, and Multichannel categories, which contain only test signals and pure tones. We uniformly sampled from each of the remaining categories. Once the annotator had annotated the example, we sampled the next example and repeated the process until we reached 500 approved files. We refer to this dataset of 500 approved files as ASFx eval and release audiocards for ASFx eval on our companion website \footnote{https://sites.google.com/view/audiocards/}.


\vspace{-0.75em}
\subsection{Generation}
\label{subsec:generation}
We prompted an LLM to generate audiocards from the available file metadata. This metadata consists of fields which are always present, such as \textit{filepath} and \textit{duration}, and other fields which are sometimes present, such as \textit{description}, \textit{keywords}, \textit{UCS category}, \textit{UCS subcategory}, \textit{extras}, and \textit{tags}.  In addition to a system prompt and warnings to stick to a JSON format, we included three handwritten audiocards in the prompt for in-context learning. We prompted the model to generate the audiocard fields in order---as the LLM is autoregressive, latter audiocard fields are conditioned on prior fields \cite{wei2022chain}. This enables us to generate audiocards grounded in audio descriptions and to generate captions of different lengths that encapsulate all relevant information about the audio file.

%



As a baseline, we prompt the LLM to sequentially generate captions in odd lengths from two words to 15 words, so that the longer captions are conditioned on the shorter captions. We use a Pixtral-12B-2409 to generate both the audiocards and the baseline captions. We henceforth refer to these as Pixtral cards and Pixtral captions.

\vspace{-0.75em}
\subsection{Improving UCS categories estimation with card classification}
\vspace{-0.5em}
\label{subsec:txtclf}

Sound effects libraries are often organized using UCS category---audio language models trained for sound design would thus benefit from learning semantic mappings between audio and UCS category. Since many examples in our data do not have human-labeled UCS tags, we aim to generate them to include in the audiocard and use in model training. However, we found in our initial experiments that LLMs used to generate audiocards could not reliably predict the UCS category and would often hallucinate, even when provided in-context examples and a list of possible UCS categories.

To this end, we trained a lightweight text classifier that takes the generated audiocards without UCS category as input, and predicts the UCS category and subcategory. We fine-tune a distilBERT-base-uncased \cite{sanh2019distilbert} with a classifier head containing two linear layers on previously generated audiocards. We train on a portion of our data with human-annotated UCS tags as targets, achieving 96\% accuracy on a held-out subset. We insert these pseudo-ground truth classifier-generated UCS predictions into the audiocards for examples without human annotated UCS tags during downstream model training.

\section{Models}
\vspace{-0.5em}
\subsection{Audio captioning}
\vspace{-0.5em}
\label{subsec:captioner}
To evaluate the usefulness of audiocards for audio captioning, we trained audio captioners to generate audiocards, called Whisper-Cards, and captions, called Whisper-Baseline, from an audio file. Whisper-Cards first predicts the signal characteristics to provide auditory grounding similar to audiocard generation in Section \ref{subsec:generation}. Specifically, we train the model to predict the audio quality, loudness, and duration, e.g., \textit{Signal characteristics -- High quality. Sound is loud. Audio is 2.98 seconds long.} Then, the captioner predicts the rest of the audiocard fields, e.g., \textit{Nouns -- Electricity, arc, spark. Verbs -- }, etc. up to the \textit{Descriptive caption} field. See Figure \ref{fig:audiocard_generation} for an example audiocard. We include additional examples of audiocards generated by the audio captioners on our companion website.

We trained Whisper-medium-v3 \cite{radford_robust_2023}, a transformer encoder-decoder sequence-sequence model pre-trained on 680k hours of weakly labeled speech data, using audio from our dataset paired with Pixtral cards or Pixtral captions for Whisper-Cards and Whisper-Baseline respectively. We downsampled all our audio files to 16kHz to match Whisper's input requirements.
We finetuned on Pixtral cards or Pixtral captions for 100k iterations using batch size 16. We used a cosine decay schedule with linear warmup for 4000 steps, and an AdamW optimizer with learning rate 1e-5. 

We evaluate our captioners using metrics commonly used in the DCASE automated audio captioning challenge -- SPIDEr \cite{liu_improved_2017} and FENSE \cite{zhou2022can}. SPIDEr is a combination of SPICE \cite{anderson_spice_2016} and CIDEr \cite{vedantam2015cider} that compares the semantic graphs of the predicted and reference captions. FENSE measures similarity using SentenceBERT \cite{reimers2019sentence} and also accounts for fluency errors.

\vspace{-0.5em}
\subsection{Text-audio contrastive representation learning}
\vspace{-0.5em}
We also evaluated the utility of audiocards for text-audio sound effects retrieval by training text-audio contrastive models. We trained a Cards-CLAP model on (audio, Pixtral audiocard field(s)) pairs by randomly selecting fields in the card during training. Through initial experiments we found the best performing model to be the one trained on a random selection from [``nouns", ``noun-verb pairs", [``nouns", ``noun-verb pairs"], [``noun-verb pairs", ``UCS Category"], ``caption in 3 words", ``caption in 7 words or less", ``descriptive caption"]. When an option with multiple fields is picked, we concatenate the two fields together. We also trained a baseline model, called Captions-CLAP, on (audio, Pixtral caption) pairs by randomly selecting different caption lengths during training. 

Our model architecture is based on the LAION-CLAP framework \cite{wu2023large}, with a RoBERTa \cite{chen_hts-at_2022} text encoder and a HTSAT \cite{liu2019roberta} audio encoder. We trained using an overall batch size 1280 for 40k steps with a peak learning rate of 1e-4, using a linear warmup of 3500 steps and a cosine decay schedule.

To evaluate text-audio retrieval, we computed audio embeddings for all examples in the dataset. Then, given a text query, we computed the cosine similarity between its embedding and all audio embeddings, picking the highest similarity top K as the retrieval results.

\section{Experiments}
\vspace{-0.5em}
\subsection{Can audio captioning models trained on audiocards generate structured metadata?}
\vspace{-0.25em}
We benchmark audiocard field generation as a structured metadata generation task for sound design using Whisper-Cards (Section \ref{subsec:captioner}), which generates audiocards from an audio file without additional prompts. As a baseline we prompt Audio Flamingo 3 (AF3) \cite{goel2025audioflamingo3}, a state-of-the-art LALM, to generate audiocard fields in the same order as Whisper-Cards, given an audio file and audiocard field descriptions. We use ASFx eval Pixtral audiocards generated from available text metadata (Section \ref{subsec:generation}) as the ground truth. We evaluate individually on the audiocard fields: \textit{nouns}, \textit{verbs}, \textit{adjectives}, \textit{noun-verb pairs}, and \textit{descriptive caption}. We also evaluate UCS category prediction as a multi-class classification task, using pseudo-ground truth from the UCS category predictor (Section \ref{subsec:txtclf}) as targets. 

\begin{figure}[t]
  \centering
  {\includegraphics[width=\columnwidth]{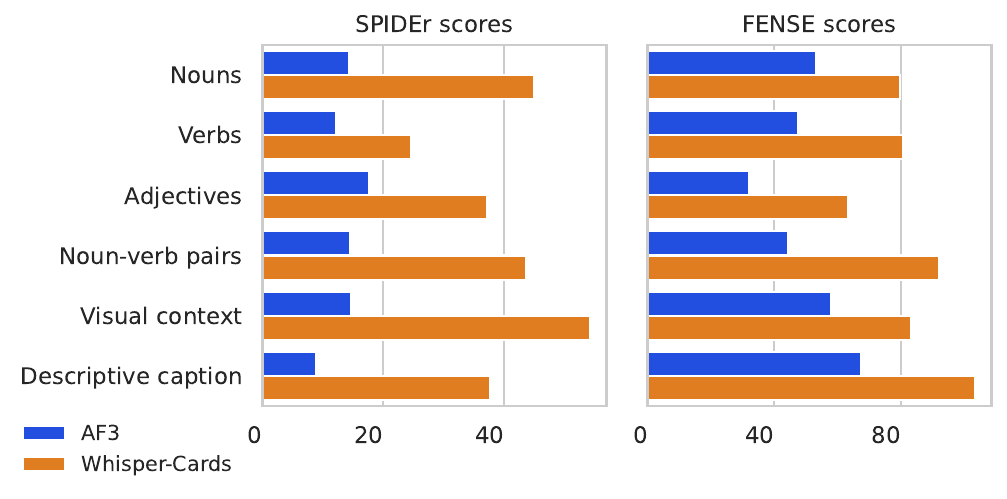}}
  \vspace{-6pt}
  \caption{Audiocard field metadata prediction evaluated on ASFx eval. We use Audio Flamingo 3 (AF3) \cite{goel2025audioflamingo3}, a state-of-the-art large audio language model, in think mode to utilize its reasoning capabilities. Whisper-Cards, trained to generate audiocards on our data, achieves superior performance on all audiocard fields.}
  \label{fig:field_prediction}
\end{figure}

 We see in Figure \ref{fig:field_prediction} that Whisper-Cards outperforms AF3 across all audiocard fields. Moreover, AF3 which is not trained using UCS tags, has multi-class UCS category prediction accuracy 0\% while Whisper-Cards achieves 31\%. While there is room for improvement, these results highlight the utility of audiocard training for sound effects metadata prediction. We also note that standard audio captioning models like Whisper-Baseline cannot generate audiocard fields as they are only trained on single sentence captions. 

\vspace{-0.75em}
\subsection{Are captions generated in audiocards better than baseline and LALM captions?}


\begin{table}[]
\caption{Audio captioning results on the Clotho and ASFx eval. Whisper-Baseline is trained on Pixtral captions, while Whisper-Cards is trained on either just the descriptive caption from Pixtral audiocards or the full audiocard. * Results from the original paper. Bold and underlined values are the best and second best models.\\}
\label{tab:cap-main}
\setlength{\tabcolsep}{1.5pt} 
\small
\centering\begin{tabular}{lcccc}
 & \multicolumn{2}{c}{\textbf{Clotho}} & \multicolumn{2}{c}{\textbf{ASFx eval}} \\
 \hline 
\textbf{Model}              & \textbf{SPIDEr} & \textbf{FENSE} & \textbf{SPIDEr} & \textbf{FENSE} \\
\hline
Whisper-Baseline                   & 21.05 & 47.61 & 7.98  & 49.78 \\
Whisper-Cards (card caption)           & 22.18 & 48.48 & \textbf{19.36} & \textbf{53.40} \\
Whisper-Cards (full card)         & 22.07 & 48.67 & \underline{18.61} & \underline{51.78} \\
\hline 
WavCaps CNN14 \cite{mei_wavcaps_2024}                & \underline{31.0}*  & -- & 3.00  & 32.20 \\
WavCaps HTSAT \cite{mei_wavcaps_2024}                & 29.7*  & -- & 3.98  & 35.27 \\
GAMA \cite{ghosh2024gama}                             & 5.34  & 43.21 & 9.30  & 45.70 \\
Audio Flamingo 2 \cite{ghosh2025audioflamingo2}                & 7.25  & 33.71 & 1.68  & 30.08 \\
Audio Flamingo 3 \cite{goel2025audioflamingo3}
& \textbf{31.88} & \textbf{52.11} & 5.27 & 32.10 \\
Audio Flamingo 3 (think) \cite{goel2025audioflamingo3}
& 13.22 & \underline{50.19} & 9.61 & 42.61 
\end{tabular}
\vspace{-1.5em}
\end{table}

To illustrate the quality of audiocard captions, we trained Whisper-Cards on just the last field from Pixtral audiocards, i.e., the descriptive caption, and Whisper-Baseline on 15 word Pixtral captions (Section \ref{subsec:captioner}). We evaluate using ASFx eval Pixtral cards and Clotho captions. In initial experiments, we found a domain gap between the training and evaluation data when models were trained on just Pixtral audiocards for our internal dataset. To rectify this, we added the Clotho development set to the training data. When training to generate the entire audiocard, we treat the Clotho caption as a \textit{descriptive caption} and ignore the other audiocard fields that don't exist in the ground truth captions in evaluation. We also compare against recent state-of-the-art LALMs, Audio Flamingo 3 (AF3) \cite{goel2025audioflamingo3} and 2 (AF2) \cite{ghosh2025audioflamingo2}, GAMA \cite{ghosh2024gama}, as well as WavCaps with CNN and HTSAT encoders \cite{mei_wavcaps_2024}. We tried several recommended prompts for AF2, AF3 and GAMA, and report metrics from the best prompt.

We report Clotho and ASFx eval metrics in Table \ref{tab:cap-main}. We find that training on Pixtral audiocards improves performance over Pixtral captions on both datasets, suggesting that audiocards contain more descriptive captions. Our models underperform AF3 and WavCaps on Clotho, likely due to the distribution mismatch between Clotho and the rest of our training data. On ASFx eval however, we find that training on audiocard captions is significantly more useful than training on baseline captions, and that Whisper-Cards significantly outperforms state-of-the-art LALMs. This suggests audiocards can significantly improve audio captioning performance on sound design specific datasets. 


%
\vspace{-0.5em}
\subsection{Does generating audiocards lead to better captions?}
We also investigate the effect of training on full audiocards on downstream captioning performance. We train Whisper-Cards to generate the entire audiocard as a series of the fields followed by their contents, given an input audio file. 
We report metrics for Whisper-Cards models trained on the full audiocard in Table \ref{tab:cap-main}. Across datasets, we find that training on the full audiocard also improves performance over Whisper-Baseline. However, we do not see a clear improvement over the Whisper-Cards model trained only on the descriptive caption. This is likely because the model can focus modeling capacity on just the descriptive caption, which is what the metrics are computed on. However, Whisper-Cards can successfully generate full audiocards while still outperforming the baseline, and is suitable for structured metadata generation to organize a sound effects library from just the raw audio files.

\vspace{-0.75em}
\subsection{Does training text-audio models on audiocards improve downstream retrieval performance?}

\begin{table}[]
\caption{Zero-shot text-audio retrieval results for models trained on audiocards (Cards-CLAP) and captions (Captions-CLAP), evaluated on Clotho and our internal dataset (ID). Audiocards outperform baseline captions across datasets, and significantly outperform comparable models on ID. * Model is trained on Clotho, result from the original paper. Bold models are best for that dataset.
}
\label{tab:retrieval-metrics}
\vspace{0.5em}
\setlength{\tabcolsep}{2.5pt} 
\small
\centering\begin{tabular}{lcccc}
\textbf{Model} & \textbf{Training regime} & \textbf{Dataset} & \multicolumn{1}{l}{\textbf{R@10}} & \multicolumn{1}{l}{\textbf{CatP@10}} \\ \hline
Captions-CLAP  & Baseline captions        & ID               & 73.45                             & 77.66                                \\ 
Cards-CLAP     & Audiocard fields         & ID               & \textbf{75.40}                    & \textbf{78.73}                       \\ 
LAION-CLAP \cite{wu2023large} & -- & ID & 24.85 & 47.10
    \\ \hline
Captions-CLAP  & Baseline captions        & Clotho           & 50.12                             & 35.00                              \\
Cards-CLAP     & Audiocard fields         & Clotho           & 52.44                    & \textbf{35.26}
\\
LAION-CLAP \cite{wu2023large} & -- & Clotho & \textbf{55.40}* & --
\end{tabular}
\vspace{-2em}
\end{table}

We evaluate retrieval results using Recall@10 on Clotho and an internal professional sound effects dataset (ID) with around 6000 samples held out during training. To quantify semantic relevance in the search results, we compute the UCS category-level P@10 as the percentage of examples in the top 10 results with the expected UCS category \cite{wilkins2023bridging}. For our internal dataset, we have ground truth UCS category annotations. For Clotho, we use the text classifier described in Section \ref{subsec:txtclf} to generate UCS category annotations. Note that neither Cards-CLAP nor Captions-CLAP are trained on Clotho or ID. 

We compare the baseline Captions-CLAP and the proposed Cards-CLAP models in Table \ref{tab:retrieval-metrics}. We find that training with audiocards leads to better downstream text-audio retrieval performance across datasets, improving recall by $4.5\%$ on Clotho and $2.6\%$ on ID. We also see an improvement in category precision, suggesting that the retrieved results are more semantically aligned with the queries. We also compare against LAION-CLAP, a popular text-audio retrieval model with a similar model architecture to Cards-CLAP \cite{wu2023large}. We find that ID is challenging for zero-shot retrieval, and that Cards-CLAP achieves comparable zero-shot performance with LAION-CLAP which is trained on Clotho.

\vspace{-1em}
\section{Conclusion}


We introduced audiocards, a novel structured metadata for sound design that are multi-field descriptions grounded in acoustic descriptors. Audiocards are more expressive than conventional single-sentence captions, enabling models to learn richer semantic associations between audio and text. Through extensive evaluations, we demonstrate that training downstream automated audio captioners and text-audio contrastive models on audiocards leads to consistent improvements over baseline captioning approaches across retrieval and captioning benchmarks. 

Beyond performance gains, our work underscores the importance of adapting audio understanding models to domain-specific requirements. To this end, we release a curated dataset of audiocards, and benchmark a novel metadata generation task on audiocard fields. Through this work, we aim to catalyze research that bridges general-purpose audio modeling with the workflows of sound designers.


\section{Acknowledgment}
\label{sec:ack}

We thank Adolfo Hernandez Santisteban for insightful discussions during this project. 
This work was partially supported by NSF awards 2504642 and 2504643.

{
\footnotesize
\bibliographystyle{IEEEtran}
\bibliography{refs,refs25,references,zotero,cam_ready}
}

\end{document}